\documentclass[aps,amsmath,amssymb,prb,superscriptaddress,twocolumn,showpacs,floatfix]{revtex4}
\usepackage{graphicx}
\usepackage[english]{babel}

\def\etcn{$\kappa$-(BE\-DT\--TTF)$_2$\-Cu$_2$(CN)$_{3}$}

\def\kcn{$\kappa$-CN}

\def\etal{{\it et al.}}
\def\vs{vs.}

\begin{document}
\title{Anisotropic charge dynamics in the quantum spin-liquid candidate $\kappa$-(BE\-DT\--TTF)$_2$\-Cu$_2$(CN)$_{3}$}
\author{M.\ Pinteri\'{c}}
\affiliation{Institut za fiziku, P.O.Box 304, HR-10001 Zagreb, Croatia}
\affiliation{Faculty of Civil Engineering, Smetanova 17, 2000 Maribor, Slovenia}
\author{M.\ \v{C}ulo}
\affiliation{Institut za fiziku, P.O.Box 304, HR-10001 Zagreb, Croatia}
\author{O.\ Milat}
\affiliation{Institut za fiziku, P.O.Box 304, HR-10001 Zagreb, Croatia}
\author{M.\ Basleti\'{c}}
\affiliation{Department of Physics, Faculty of Science, University of Zagreb, P.O.Box331, HR-10001 Zagreb, Croatia}
\author{B.\ Korin-Hamzi\'{c}}
\affiliation{Institut za fiziku, P.O.Box 304, HR-10001 Zagreb, Croatia}
\author{E.\ Tafra}
\affiliation{Department of Physics, Faculty of Science, University of Zagreb, P.O.Box331, HR-10001 Zagreb, Croatia}
\author{A.\ Hamzi\'{c}}
\affiliation{Department of Physics, Faculty of Science, University of Zagreb, P.O.Box331, HR-10001 Zagreb, Croatia}
\author{T.\ Ivek}
\affiliation{Institut za fiziku, P.O.Box 304, HR-10001 Zagreb, Croatia}
\affiliation{1.\ Physikalisches Institut, Universit\"{a}t Stuttgart, Pfaffenwaldring 57, D-70550 Stuttgart, Germany}
\author{T.\ Peterseim}
\affiliation{1.\ Physikalisches Institut, Universit\"{a}t Stuttgart, Pfaffenwaldring 57, D-70550 Stuttgart, Germany}
\author{K.\ Miyagawa}
\affiliation{Department of Applied Physics, University of Tokyo, Tokyo 113-8656, Japan}
\author{K.\ Kanoda}
\affiliation{Department of Applied Physics, University of Tokyo, Tokyo 113-8656, Japan}
\author{J.\ A.\ Schlueter }
\affiliation{Material Science Division, Argonne National Laboratory - Argonne, Illinois 60439-4831, U.S.A.}
\author{M.\ Dressel}
\affiliation{1.\ Physikalisches Institut, Universit\"{a}t Stuttgart, Pfaffenwaldring 57, D-70550 Stuttgart, Germany}
\author{S.\ Tomi\'{c}}
\affiliation{Institut za fiziku, P.O.Box 304, HR-10001 Zagreb, Croatia}
\email{stomic@ifs.hr}
\homepage{http://real-science.ifs.hr/}

\date{\today}

\begin{abstract}
We have in detail characterized the anisotropic charge response of the
dimer Mott insulator $\kappa$-(BEDT-TTF)$_2$\-Cu$_2$(CN)$_3$ by dc conductivity,
Hall effect and dielectric spectroscopy.
At room temperature the Hall coefficient is positive and close to the value
expected from stoichiometry; the temperature behavior follows the dc resistivity $\rho(T)$.
Within the planes the dc conductivity is well described by
variable-range hopping in two dimensions;
this model, however, fails for the out-of-plane direction.
An unusually broad in-plane dielectric relaxation is detected below about 60\,K;
it slows down much faster than the dc conductivity following an Arrhenius law.
At around 17\,K we can identify a pronounced dielectric anomaly concomitantly with anomalous features in the mean relaxation time and spectral broadening.
The out-of-plane relaxation, on the other hand, shows a much weaker dielectric anomaly;
it closely follows the temperature behavior of the respective dc resistivity.
At lower temperatures, the dielectric constant becomes smaller
both within and perpendicular to the planes; also the relaxation levels off.
The observed behavior bears features of relaxor-like ferroelectricity.
Because heterogeneities impede its long-range development,
only a weak tunneling-like dynamics persists at low temperatures.
We suggest that the random potential and domain structure gradually emerge
due to the coupling to the anion network.
\end{abstract}


\maketitle

\section{Introduction}
\label{sec:Intro}
Strongly correlated systems show a number of anomalous features and exotic quantum phases
that cannot be understood within the standard weakly-interacting quasi-particle picture of condensed-matter physics. One example is the quantum spin liquid (QSL), first conceived by Anderson,\cite{Anderson1973} and eventually observed in systems with strong antiferromagnetic interactions in the vicinity of the Mott transition.
It is characterized by the absence of magnetic order
due to the interplay of quantum effects and frustration occurring on either kagome or triangular lattice.\cite{BalentsNature2010}
Many types of QSL states are predicted theoretically and are characterized by a variety of magnetic excitation spectra with or without an energy gap.
The  hallmark for most of them is
that instead of conventional $S=1$ spin-wave excitations, so-called magnons,
fractionalized $S=1/2$ charge-neutral excitations occur, which are
named spinons.\cite{LeePRL2005}
They originate in the separation of charge and spin degrees of freedom
near the Mott transition.
Common to all QSL states these spinons are always coupled to internal gauge fields yielding a number of exotic properties.\cite{LeeScience2008}
The nature of spinons is not quite clear in two dimensions. For the kagome lattice
it was recently shown that spinons can be considered as domain walls
free to move, similar to the one-dimensional case.\cite{YanScience2011}
Although the QSL concept has been proposed a long time ago,\cite{Anderson1973} only recent experiments backed up with novel advances in theory could provide evidence
for its realization in actual materials.
These include triangular organic systems \etcn{}, EtMe$_3$Sb[Pd(dmit)$_2$]$_2$, $\kappa$-H$_3$(Cat-EDT-TTF)$_2$
and the kagome-lattice system ZnCu$_3$(OH)$_6$Cl$_2$.\cite{ShimizuPRL2003,YamashitaKatoScience2010,YamashitaKatoNatureComm2011, IsonoPRL2014,HeltonPRL2007}

Most prominent is the organic dimer Mott insulator \etcn{} [hereafter labeled \kcn{},
BEDT-TTF stands for bis-(ethyl\-ene\-di\-thio)\-te\-tra\-thia\-ful\-va\-lene] which was suggested
as a two-dimensional spin-liquid compound already a decade ago.\cite{ShimizuPRL2003} Neither magnetic order nor structural distortions have been detected down to 30\,mK, while at the same time the susceptibility behavior indicates strong antiferromagnetic (AFM) exchange coupling $J\approx 250$\,K according to the $S =1/2$ triangular-lattice Heisenberg model. The residual spin susceptibility and power-law temperature dependence of the NMR spin-lattice relaxation rate
indicate low-lying spin excitations. Since the on-site Coulomb repulsion $U$ is comparable to the bandwidth $U/W \approx 1.8$, where $W = 4t$ and $t$ is the hopping integral,\cite{KandpalPRL2009}, this system lies near the Mott transition.
Indeed, a pressure of about 4\,kbar is sufficient to suppress the insulating and establish a metallic and superconducting state.\cite{KurosakiPRL2005}

\begin{figure}
\centering
\includegraphics[clip=true,width=\columnwidth]{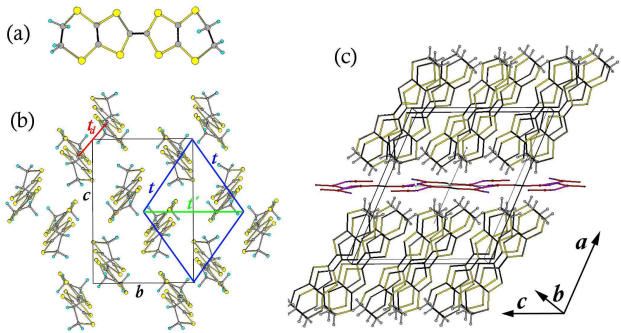}
\caption{(Color online) (a) Schematic drawing of a BEDT-TTF molecule: yellow and gray symbols denote sulfur and carbon atoms. The hydrogen atoms denoted by cyan symbols belong to the ethylene CH$_2$ side-groups which are slightly twisted out of plane.
(b) View of BEDT-TTF dimers in the $bc$ plane projected along the $a$-axis; an almost isotropic triangular lattice is denoted by full thick lines; the interdimer transfer integrals are labeled by $t$ and $t^{\prime}$, while the intradimer transfer integral is labeled by $t_d$; the unit cell is denoted as a rectangle. (c) Side view of extended unit cell of $\kappa$-(BE\-DT\--TTF)$_2$\-Cu$_2$(CN)$_{3}$.}
\label{fig:structurecomposite}
\end{figure}

Crystallographic considerations suggest that frustration may have geometric origin.
The compound consists of BEDT-TTF layers in crystallographic $bc$ plane
separated by sheets of interconnected anions  [Fig.\ \ref{fig:structurecomposite}(c)].\cite{Geiser1991,JeschkePRL2012}
Each unit cell contains four strongly dimerized BEDT-TTF molecules with each dimer oriented approximately perpendicularly to its neighbors as sketched in Fig.\ \ref{fig:structurecomposite}(b).
Half an electron per BEDT-TTF is transferred to the anion layer, leaving
behind one hole per dimer.
The two-dimensional mesh of interdimer transfer integrals connects the spin-1/2
in an almost isotropic triangular lattice. From density functional theory calculations we know that at room temperature
the ratio of inter-dimer transfer integrals $t^{\prime}/t=0.83$ are very close to unity. \cite{KandpalPRL2009,Nakamura09,JeschkePRL2012}
Hence, the system is described as a half-filled band with strong spin frustration
that prohibits magnetic ordering and suggests a quantum spin-liquid state.
In spite of a great deal of experimental and theoretical studies,
no consistent picture of the spin-liquid state can be drawn and
no definite understanding has been reached yet on the ground state in \kcn{}.
The open questions mainly concern the existence of a spin gap, the nature of the charge-spin coupling
and the origin of charge and dielectric response observed at audio, terahertz and optical frequencies.

On the one hand, the $T$-linear contribution to the heat capacity implies the presence of gapless low-lying spinon excitations,\cite{ShimizuPRL2003, YamashitaKanoda2008} but on the other hand thermal conductivity measurements indicate that if those spin excitations are present they experience some instability leading to
a small gap of about 0.5\,K.\cite{YamashitaMatsuda2009}
This apparent contradiction can be reconciled taking into account a $^{13}$C-NMR spectral line broadening emerging without AFM critical fluctuations. These results indicate a spatially nonuniform staggered magnetization induced in the spin liquid by applied magnetic field.\cite{Kawamoto2004, Shimizu2006}
Below 100\,K numerous physical properties in both the spin and charge sectors
evidence some anomalous behavior that infers an exotic charge-spin coupling and the possibility
that it may play a pertinent role in the formation of the QSL at low temperatures.
As first noted by Abdel-Jawad {\it et al.}\cite{Abdel-Jawad2010} the out-of-plane relaxor-like dielectric response in the audio-frequency range becomes rather strong below about 60\,K
although the dimers do not contain any static electric dipoles and
any sort of charge-ordering can be ruled out down to low temperatures.\cite{Sedlmeier2012}
Nevertheless, the dielectric constant grows on cooling and seems to freeze at about 6\,K;
at this temperature also the microwave dielectric constant shows the strongest change.\cite{Poirier2012}
A broad band was observed in the optical conductivity around 1\,THz which grows below 60\,K;
it was attributed to the collective excitation of the fluctuating intradimer electric dipole.\cite{ItohPRL2013}
In the same temperature range the spin susceptibility decreases gradually
and  $(T_1T)^{-1}$ obtained from $^{13}$C-NMR measurements increases,
both indicating the development of AFM correlations.
Eventually the susceptibility drops sharply below 10\,K where $(T_1T)^{-1}$ shows a broad peak.\cite{Kawamoto2004}
The latter, although reminiscent of the AFM critical fluctuations, could not be associated to an AFM ordering
since no additional internal field was detected.
The NMR spin-lattice relaxation rate $(T_1)^{-1}$ gradually decreases on cooling;
however, at 6\,K the behavior changes: a dip-like structure is followed by a broad peak around 1\,K.\cite{ShimizuPRL2003}
The crossover temperature of 6\,K also becomes evident as a broad hump structures in heat capacity and
thermal conductivity,\cite{YamashitaKanoda2008, YamashitaMatsuda2009} by anisotropic lattice effects\cite{MannaPRL2010} and
by an anomaly in the out-of-plane phonon velocity and ultrasound attenuation;\cite{Poirier2013}
the two latter results indicate the involvement of the spin-phonon coupling.

There were several theoretical approaches to describe the spin-charge coupling in \kcn{} and to explain the experimentally observed dielectric response. Hotta \cite{Hotta} suggested that quantum electric dipoles on dimers interact with each other via dipolar-spin coupling; a similar idea was put forward by Naka and Ishihara. \cite{Naka}
Clay, Mazumdar and collaborators \cite{Sumit} suggested the formation of a paired electron crystal
implying frustration-induced charge disproportionation.
Since recent spectroscopic evidence discarded any static electric dipoles on the dimers,\cite{Sedlmeier2012}
these and other models \cite{Gomi13} were put into question; different approaches are required for a satisfactory theoretical description.
Qi \etal{}\cite{QiPRB2008} proposed that the 6\,K phase transition denotes a change in the charge sector:
it is triggered by the formation of an excitonic insulator within the $Z_2$ spin liquid state with a tiny spin gap.
Due to the proximity of the quantum phase transition to a magnetically ordered state the gap is rather small.
This suggestion is under discussion because the lattice symmetry should be reduced by such a transition
which has not been observed yet.
Yamamoto {\it et al.}\cite{Yamamoto08,Yamamoto10a,Yamamoto10b} reported that below the charge-order transition
a significant second harmonic signal develops in $\alpha$-(BEDT-TTF)$_2$I$_3$ and $\alpha^\prime$-(BEDT-TTF)$_2$IBr$_2$,
which is a strong proof of symmetry breaking. Second harmonic generation could not be achieved in $\kappa$-phase BEDT-TTF.

Another interesting, but yet unsolved issue is the charge response in the infrared spectral range.
Based on $U(1)$ gauge field it was predicted that spinons give a power-law contribution to
the optical conductivity within the charge gap.\cite{NgPRL2007}
Although an excess conductivity with power-law behavior was detected, the observed features do not agree quantitatively
with the theoretical predictions:
the power-law exponent was significantly smaller and
there was no increase in the slope when going from low to high frequencies the way it is predicted.\cite{Elsaesser2012}

In order to clarify the nature of the spin-liquid state in $\kappa$-(BE\-DT\--TTF)$_2$\-Cu$_2$(CN)$_{3}$
and in particular the charge dynamics, we have conducted a detailed characterization of the in-plane and out-of plane
dc transport and investigated the dielectric response in the audio frequency range.
Most importantly we found a close relation between the temperature behavior of both phenomena:
while the temperature dependence of the in-plane dc conductivity indicates variable-range hopping in two dimensions,
the dielectric response reveals signatures of relaxor ferroelectricity.
Since the latter requires some heterogeneity, we suggest that it is caused by
the inversion symmetry in the anion network being broken on a local scale
so that the average structure with $P2_1/c$ symmetry can be retained on a global scale. Our structural refinements verify this, showing that both the $P2_1/c$, and the $P2_1$ symmetry without the inversion center characterize the crystal structure with comparably high agreement factors.
We propose that the local rearrangement in the anions is transferred to the BEDT-TTF molecules via hydrogen bonds,
yielding the inhomogeneity and a random domain structure throughout the whole crystal.
Charge defects generated in interfaces respond to an applied ac electric field and
are thus responsible for the slow relaxation observed in the audio spectrum.
In order to verify our hypothesis, we have conducted investigations on single crystals of three completely different
origin and distinct differences in the degree of disorder.

\section{Samples and Methods}
For the experiments we selected high-quality single crystal of \etcn\ grown at Argonne and Tokyo
by different conditions within the standard synthesis procedures;\cite{Geiser1991,note1} the
three batches harvested for this study are labelled by sA, sB and sC.
While all specimens exhibit qualitatively a similar behavior, quantitative differences can be identified
in the parameters characterizing dc and ac transport properties.
Similar observations were previously reported by the microwave and thermal conductivity measurements.\cite{Poirier2012,YamashitaMatsuda2009}
Confining ourselves to slow cooling rates (2--15\,K/h)
the data are not influenced by different cycling routes. 

The typical crystal dimensions are $1.5 \times 0.5 \times 0.05$\,mm$^3$.
We performed electrical measurements along the three crystallographic axes:
$c$-axis and $b$-axis within molecular planes and $a^\ast$-axis perpendicular to the ($bc$) plane.
The single crystals were oriented on the basis of mid-infrared spectra recorded before and x-ray back-reflection
Laue photographs done after the electrical experiments.

DC resistivity was measured between room temperature and 10\,K by standard four-contact technique (see inset of Fig.\ \ref{fig:samplegeometryandresistivity}). \cite{Basletic2014} 
The Hall effect was obtained with the magnetic field $B\leq 5$\,T along the $a^\ast$-axis and
the current along the $c$-axis ($\mathbf{E}\parallel c$) for the majority of samples.
Dielectric spectroscopy measurements were performed in the frequency range 40\,Hz--10\,MHz between 300 and 10\,K.
To obtain spectra of the complex dielectric function we measured the complex conductance
using an Agilent 4294A and HP 4284A precision impedance analyzers.
The employed ac signal levels of 10\,mV and 50\,mV were confirmed to be well within the linear response regime.
Best contacts for transport and dielectric measurements are produced by carbon paint directly applied to the sample surface.
Extrinsic effects, especially those due to contact resistance and surface layer capacitance, were ruled out with scrutiny.\cite{IvekPRB2011} For this reason, all data from dielectric measurements obtained above 60\,K were discarded.
In addition, measurements on samples with contacts made on pre-evaporated gold pads were fully discarded because they showed significant nonlinear effects due to large contact capacitances throughout the whole measured temperature range.

Finally, in order to examine the crystal symmetry at the local scale the x-ray single crystal data were collected at 300\,K and at 100\,K with graphite monochromated Mo-Kα radiation by $\omega$-scans on an Oxford Diffraction KM4 XCALIBUR2 CCD diffractometer. Data acquisition and reduction was performed using the CrysAlisPro software package (Version 1.171.37.33). The structure was solved by direct methods using SHELXS. \cite{Sheldrick2008} The refinement procedure was performed by the full-matrix least-squares method based on F-square against all reflections using SHELXL. 

\section{Results and Analysis}
\subsection{DC transport and Hall coefficient}
\begin{figure}[b]
\centering
\includegraphics*[width=7cm]{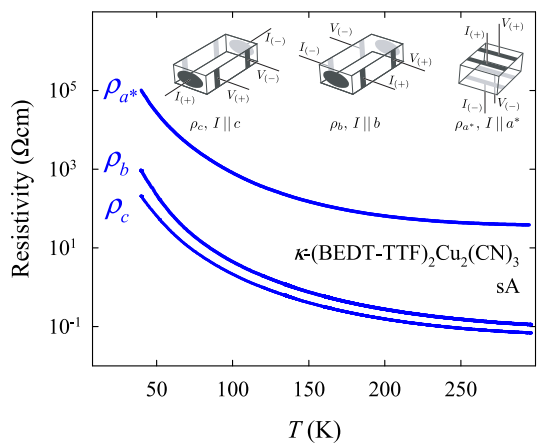}
\caption{(Color online) Temperature dependence of the resistivities $\rho_c$, $\rho_b$ and $\rho_{a^\ast}$
for representative  $\kappa$-(BE\-DT\--TTF)$_2$\-Cu$_2$(CN)$_{3}$ single crystals sA. The insets explain the sample geometry and contact arrangements utilized for electrical resistivity measurements. Here $I$ denotes applied electrical current, and $V$ measured voltage.}
\label{fig:samplegeometryandresistivity}
\end{figure}

The temperature-dependent electrical resistivity exhibits an insulating behavior for all three crystallographic directions.
Data obtained on single crystals from the sA synthesis are displayed in Fig.\ \ref{fig:samplegeometryandresistivity} as an example.

At room temperature the in-plane dc resistivity reveals an extremely small anisotropy of $\rho_b/\rho_c \approx 1.5$, whereas the out-of-plane anisotropy is in the range of 100 to 1000. The values of anisotropy weakly change with decreasing temperature, but for $T<50$\,K the change becomes more pronounced (not shown). 

\begin{figure}[h]
\includegraphics[clip,width=0.8\columnwidth]{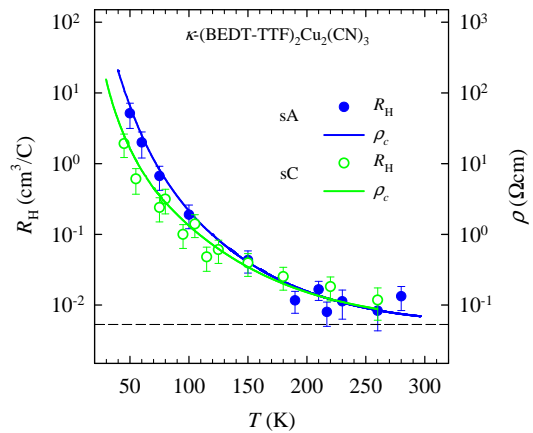}
\caption{(Color online) Comparison of the Hall coefficient $R_H(T)$ for $\mathbf{E}\parallel c$ and $\mathbf{B}\parallel a^\ast$ (symbols) and the dc resistivity $\rho(T)$ for $\mathbf{E}\parallel c$ (full line) for $\kappa$-(BE\-DT\--TTF)$_2$\-Cu$_2$(CN)$_{3}$ representative single crystals from sA (full blue symbols) and sC (empty green symbols). The dashed lines denotes the value of Hall coefficient expected for two holes per two dimers in the unit cell at room temperature.}
\label{fig:HallAC}
\end{figure}

Figure \ref{fig:HallAC} shows the Hall coefficient $R_H$ as a function of temperature for two single crystals syntheses sA and sC, respectively. $R_H$ is positive and
at $T=300$\,K it agrees with the value expected for two holes per two dimers in the unit cell; it closely follows the temperature behavior of the dc resistivity. Upon cooling down to 50\,K, $R_H(T)$ increases by three orders of magnitude, indicating the complete freezing out of mobile carriers.

\begin{figure}
\includegraphics[clip,width=0.8\columnwidth]{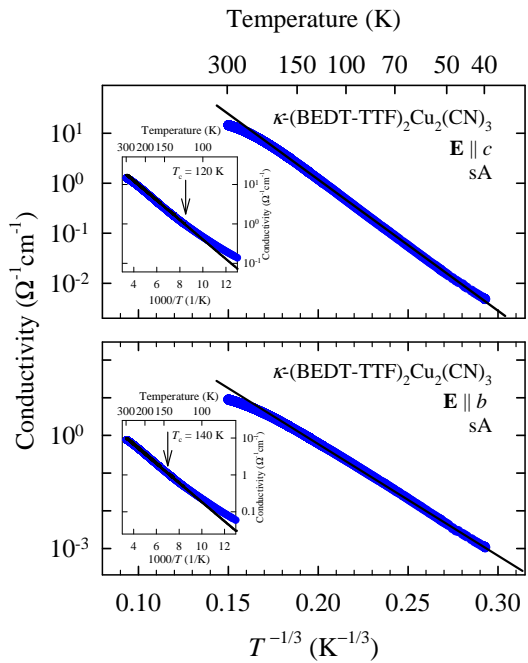}
\caption{(Color online) DC conductivity of a $\kappa$-(BE\-DT\--TTF)$_2$\-Cu$_2$(CN)$_{3}$ representative single crystal from synthesis sA for $\mathbf{E}\parallel c$ (upper panel) and for $\mathbf{E}\parallel b$ (lower panel) as a function of $T^{-1/3}$. The conductivity follows a simple activation above $T_c$ (full line in the inset) indicating nearest-neighbor hopping. Below $T_c$, the observed behavior corresponds to the regime of variable-range hopping in two dimensions, $\sigma(T) \simeq \exp[-(T_0/T)^{1/3}]$ (see text). Qualitatively similar observations have been made for crystals of batch sB and sC.}
\label{fig:VRHA}
\end{figure}

As demonstrated in the insets of Fig.\ \ref{fig:VRHA}, only above $T_c \approx 130$\,K
the in-plane dc conductivity along the $c$- and $b$-axes
can be fitted by the activated behavior expected for an insulator with the carriers contributing to
the conductivity by the nearest-neighbor hopping:
\begin{equation}
\sigma(T) \propto \exp(-\Delta/T)
\end{equation}
with $\Delta \approx 500$\,K. At lower temperatures ($T<T_c$) the charge transport can be perfectly presented in a
$\log[\sigma(T)]$ \vs\ $T^{-1/3}$ plot displayed in the main panels of Fig.\ \ref{fig:VRHA} for the
two directions $\mathbf{E}\parallel c$ and $\mathbf{E}\parallel b$.
This evidences a variable-range hopping (VRH) behavior in two dimensions ($d=2$)
\begin{equation}
\sigma(T) \propto \exp\left[-(T_0/T)^{1/(d+1)}\right],
\label{eq:VRH}
\end{equation}
where $T_0$ is the corresponding  activation energy.\cite{Mott1971}
These results clearly demonstrate that the hopping mechanism is responsible for the charge transport within the molecular planes.
Qualitatively similar results are observed in all crystals of the three different batches, sA, sB and sC.

\begin{table} 
\centering
\caption{In-plane dc transport parameters of $\kappa$-(BE\-DT\--TTF)$_2$\-Cu$_2$(CN)$_{3}$ for representative single crystals from three different syntheses sA, sB and sC for $\mathbf{E}\parallel c$ and for $\mathbf{E}\parallel b$.}
\begin{tabular*}{\linewidth}{@{\extracolsep{\fill}}ccccc}
\hline\hline
Axis &  & sA & sB & sC \\
\hline
$c$-axis & $\Delta$ (K) & $530 \pm 20$ & $500 \pm 20$ & $480 \pm 20$ \\
$c$-axis & $T_c$ (K) & $120 \pm 10$ & $130 \pm 10$ & $130 \pm 10$ \\
$c$-axis & $T_0$ (eV) & 23 & 17.5 & 9 \\
$b$-axis & $\Delta$ (K) & $600 \pm 20$ & $600 \pm 20$ & $600 \pm 20$ \\
$b$-axis & $T_c$ (K) & $140 \pm 10$ & $130 \pm 10$ & $120 \pm 10$ \\
$b$-axis & $T_0$ (eV) & 31 & 20 & 5.7 \\
\hline\hline
\end{tabular*}
\label{tab:parameters}
\end{table}

The nearest-neighbor hopping dominates the transport above  $T_c \approx 130$\,K, where it crosses over to variable-range hopping at low temperatures according to Eq.\ (\ref{eq:VRH}) (the corresponding values are listed in Table \ref{tab:parameters}). The activation energies $\Delta$  for the nearest-neighbor hopping and the crossover temperatures $T_c$ into the VRH regime are similar along both in-plane directions. The same holds for the VRH parameter $T_0$, which presents somewhat smaller values along the $c$-axis than along the $b$-axis; this agrees with the higher conducting direction.
For all three directions $\Delta$ is of the order of several hundreds kelvins
for the complete temperature range from 100 to 300\,K; this clearly indicates
that no true charge gap opens, in agreement with previous optical results.\cite{Kezsmarki2006,Elsaesser2012}
At the quantitative level, it should be noted that $T_0$ values are largest for samples from sA and smallest for sC crystals in accord with the VRH mechanism: for better conducting samples a lower $T_0$
is expected. The sA samples rise steeper in $\rho(T)$ when cooled below about 150\,K than the crystals from the sB and sC synthesis, as displayed in Fig.\ \ref{fig:HallAC} (the data for the sB sample are omitted for clarity); this tendency is observed for all three crystallographic directions.
It is interesting to note that VRH fits to the $\sigma(T)$ along the $a^\ast$-axis are not satisfactory and fail to give a meaningful description: although the conductivities along the $a^\ast$-axis are more than 100 times smaller than the in-plane conductivities, the respective values of $T_0$ are similar. In addition, VRH fits for $d=1$ and 3 also fail and lead to unreasonably low and high $T_0$ values, respectively.
Collecting all these pieces of evidence leads to the conclusion that going from sA, via sB to sC samples of \kcn{} disorder increases.

\subsection{Dielectric response}
\subsubsection{Frequency and temperature dependences}
\label{sec:frequencytemperaturedependence}
\begin{figure}
\includegraphics[clip,width=0.7\columnwidth]{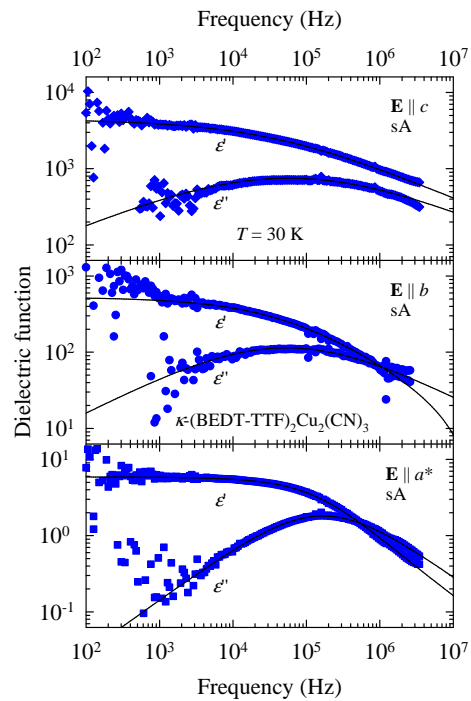} 
\caption{(Color online) Double logarithmic plot of the frequency
dependence of the real ($\varepsilon'$) and imaginary ($\varepsilon''$) parts of
the dielectric function in representative single crystals of $\kappa$-(BE\-DT\--TTF)$_2$\-Cu$_2$(CN)$_{3}$ from synthesis sA. The $T=30$\,K data are plotted for $\mathbf{E}\parallel c$ (upper panel), for $\mathbf{E}\parallel b$ (middle panel) and for $\mathbf{E}\parallel a^\ast$ (lower panel). The full lines are fits to a generalized Debye function Eq.\ (\ref{eq:gDebye}), as discussed in the text.}
\label{fig:SpectraA}
\end{figure}

\begin{figure}
\includegraphics[clip,width=0.8\columnwidth]{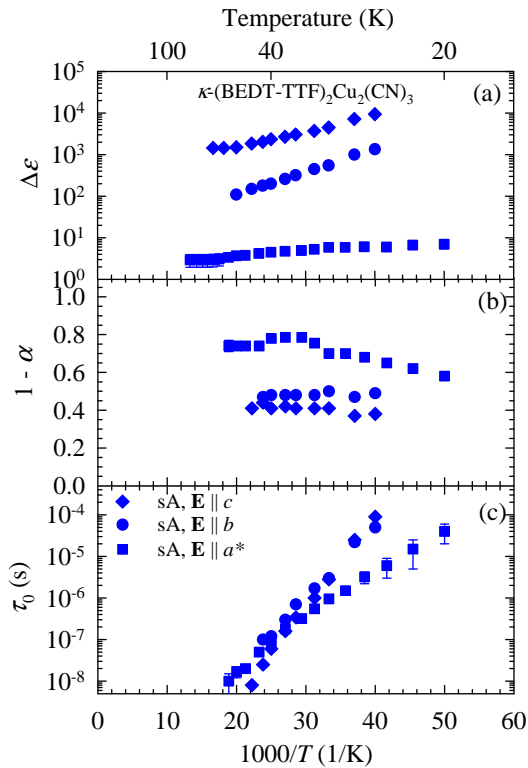} 
\caption{(Color online) Dielectric strength (upper panel), distribution of relaxation times (middle panel) and mean relaxation time (lower panel) in sA representative single crystals of $\kappa$-(BE\-DT\--TTF)$_2$\-Cu$_2$(CN)$_{3}$ as a function of inverse temperature. Diamonds, circles and squares are for $\mathbf{E}\parallel c$, $\mathbf{E}\parallel b$ and $\mathbf{E}\parallel a^\ast$, respectively. Above 50\,K the mode is outside the frequency window and only the dielectric strength can be determined.}
\label{fig:3panel3axesA}
\end{figure}

In Fig.\ \ref{fig:SpectraA} we exhibit typical spectra of the real and imaginary parts of the dielectric function $\varepsilon=\varepsilon'-{\rm i} \varepsilon''$ for sA crystals of \kcn{} recorded at $T=30$\,K
for the directions $\mathbf{E}\parallel c$, $\mathbf{E}\parallel b$ and $\mathbf{E}\parallel a^\ast$.
These spectra bear features commonly found in a relaxation-type of dielectric response.
A low-frequency constant plateau is followed by a decrease of $\varepsilon^\prime$ with frequency; associated with this drop is a Kramers-Kronig-consistent increase in $\varepsilon^{\prime\prime}$ which gives it a characteristic bell-like peak in a log-log presentation. The main features of this relaxation are well described by the generalized Debye expression:
\begin{equation}
\varepsilon(\omega)-\varepsilon_{\mathrm{HF}} =
	\frac{\Delta\varepsilon}{1+({\rm i}  \omega \tau_0)^{1-\alpha}} \quad,
\label{eq:gDebye}
\end{equation}
where $\Delta\varepsilon = \varepsilon_0 - \varepsilon_{\mathrm{HF}}$ corresponds to the strength of the mode; $\varepsilon_0$ and $\varepsilon_{\mathrm{HF}}$ are the static and high-frequency dielectric constant, respectively; $\tau_0$ is the mean relaxation time; and $1-\alpha$ is the symmetric broadening of the relaxation time distribution function. The temperature dependence of the extracted parameters $\Delta\varepsilon$, $1-\alpha$ and $\tau_0$ is plotted in Fig.\ \ref{fig:3panel3axesA} as a function of $1/T$.\cite{note2}
Compared to similar systems \cite{IvekPRB2011} the relaxation
appears rather broad; the strength of the dielectric response within planes ($\mathbf{E}\parallel b$, $\mathbf{E}\parallel c$) is on the order of $100-1000$, while the dielectric strength perpendicular to planes ($\mathbf{E}\parallel a^\ast$) is only of the order of 10 and less.

\begin{figure}
\includegraphics[clip,width=0.8\columnwidth]{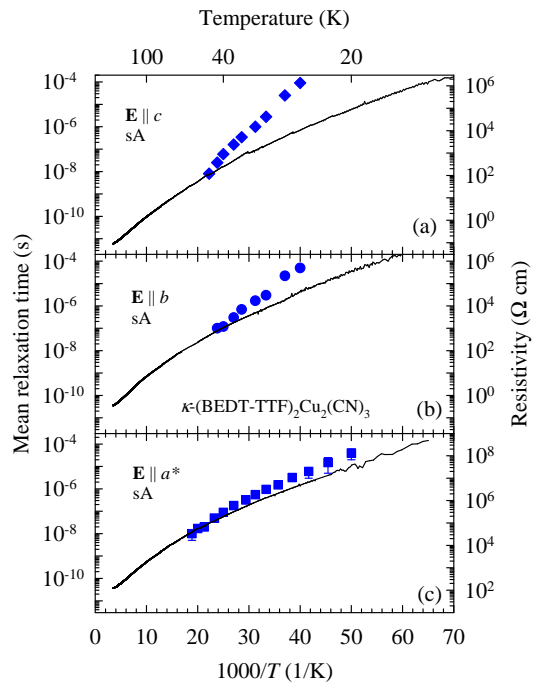} 
\caption{(Color online) Comparing the temperature behavior of the mean relaxation time (symbols, left axis) and dc resistivity (full lines, right axis).
The data taken at $\kappa$-(BE\-DT\--TTF)$_2$\-Cu$_2$(CN)$_{3}$ representative single crystals of synthesis  sA are plotted as a function of inverse temperature. Diamonds, circles and squares correspond to (a) $\mathbf{E}\parallel c$, (b) $\mathbf{E}\parallel b$, and (c) $\mathbf{E}\parallel a^\ast$, respectively.}
\label{fig:tau3axesA}
\end{figure}

Let us now consider the temperature dependence of the mean relaxation time plotted in Fig.\ \ref{fig:tau3axesA}.
The most important point is the strong deviation of $\tau_0(T)$ from the behavior of the corresponding dc resistivity.
This is most pronounced for the in-plane relaxation along the $c$-axis, but also significant for ${\bf E}\parallel b$;
a similar deviation along the $a^\ast$-axis is much less pronounced.

The second important observation is the anomalously broad in-plane loss curve with $1-\alpha \approx$ 0.45; for the perpendicular direction the loss curve is rather narrow at high temperatures ($1-\alpha \approx $ 0.75) but broadens significantly when the temperature decreases so that $1-\alpha \approx  0.55$ at $T=20$\,K. Both features are considered fingerprints of cooperative motion and glassy freezing in relaxor ferroelectrics.\cite{Cross1987}

For a quantitative analysis, we have fitted our data by the expression
\begin{equation}
\tau_0 = \tau_{00} \left( T/T_0 - 1 \right)^{-z\nu}
\label{eq:CSD}
\end{equation}
commonly used to describe the critical slowing down, and by the Vogel-Fulcher expression
\begin{equation}
\tau_0 = \tau_{00} \exp\left[\Delta_g/(T-T_c)\right]
\label{eq:VF}
\end{equation}
as the well-known description of the relaxation time $\tau_0$ in glassy systems.
$\Delta_g$ is an effective energy barrier, and $T_0$ and $T_c$ are temperatures where the systems's mean relaxation time diverges.
Following the approach by Abdel-Jawad \etal{} \cite{Abdel-Jawad2010} to describe the perpendicular response ($\mathbf{E}\parallel a^\ast$) by keeping $T_c = 6$\,K fixed, we were able to perform satisfactory fits to all three directions using the Vogel-Fulcher expression. The results summarized in Table \ref{tab:VF} are similar to the ones found previously. It is not  surprising, that the Vogel-Fulcher  expression gives even better fits when all three parameters are left free (not shown); however, the obtained negative critical temperatures $T_c$ and much larger $\Delta_g$ challenge the application of the Vogel-Fulcher formula on our data. Moreover, we suspect that an even broader temperature range is needed in order to clearly discern
a critical slowing down type of behavior as described by Eq.\ (\ref{eq:CSD}) and Eq.\ (\ref{eq:VF}).

\begin{table} [h]
\centering
\caption{Vogel-Fulcher parameters (VF) of the mean relaxation time $\tau_0$ for representative single crystals from sA and sB synthesis of $\kappa$-(BE\-DT\--TTF)$_2$\-Cu$_2$(CN)$_{3}$ for $\mathbf{E}\parallel c$, $\mathbf{E}\parallel b$ and $\mathbf{E}\parallel a^\ast$. Note that for the fit $T_c$ is fixed to 6\,K.}
\begin{tabular*}{\linewidth}{@{\extracolsep{\fill}}ccccc}
\hline\hline
Axis & VF & sA & sB \\
\hline
$c$-axis & $\Delta_g$ (K)	& $340 \pm 10$ 	& $200 \pm 10$ \\
$c$-axis & $T_c$ (K) 		& $6$   & $6$ \\
$b$-axis & $\Delta_g$ (K)	& $260 \pm 10$ 	& $120 \pm 50$ \\
$b$-axis & $T_c$ (K) 		& $6$   & $6$ \\
$a^\ast$-axis & $\Delta_g$ (K)		& $166 \pm 8$ & $178 \pm 5$ \\
$a^\ast$-axis & $T_c$ (K) 			& $6$   & $6$ \\
\hline\hline
\end{tabular*}
\label{tab:VF}
\end{table}

Thus, we concluded that the fits to the Arrhenius type of gradual slowing down 
\begin{equation}
\tau_0 = \tau_{00} \exp\left(\Delta/T\right)
\label{eq:Arrhenius}
\end{equation}
are more appropriate to describe our dielectric relaxation data; here we follow the common definition of a glass transition temperature $T_g$ as the temperature where the relaxation time $\tau_0$ extrapolates to the value of 100\,s. The obtained values for $T_g$ are between 10 and 15\,K (Table \ref{tab:Arhenius}), exactly within temperature range where the anomalies of a number of physical properties are observed, as summarized in Sec.\ \ref{sec:Intro} above.

\begin{table} 
\centering
\caption{Arrhenius activation energy $\Delta$ of mean relaxation time $\tau_0$ and glass temperatures $T_g$ at $\tau_0 = 100$\,s for representative single crystals from sA and sB synthesis of $\kappa$-(BE\-DT\--TTF)$_2$\-Cu$_2$(CN)$_{3}$ for $\mathbf{E}\parallel c$, $\mathbf{E}\parallel b$ and $\mathbf{E}\parallel a^\ast$.}
\begin{tabular*}{\linewidth}{@{\extracolsep{\fill}}ccccc}
\hline\hline
Axis & Arrhenius & sA & sB \\
\hline
$c$-axis & $\Delta$ (K)		& $510 \pm 10$ & $330 \pm 20$ \\
$c$-axis & $T_g$ (K) 	& $14.9$ & $10.6$ \\
$b$-axis & $\Delta$ (K)		& $392 \pm 14$ & $212 \pm 5$ \\
$b$-axis & $T_g$ (K) 	& $13.0$ & $7.5$ \\
$a^\ast$-axis & $\Delta$ (K)		& $265 \pm 8$ & $319 \pm 6$ \\
$a^\ast$-axis & $T_g$ (K) 	& $9.6$ & $10.3$ \\
\hline\hline
\end{tabular*}
\label{tab:Arhenius}
\end{table}

\subsubsection{Dependence on disorder}
\label{sec:disorder}
\begin{figure}
\includegraphics[clip,width=0.8\columnwidth]{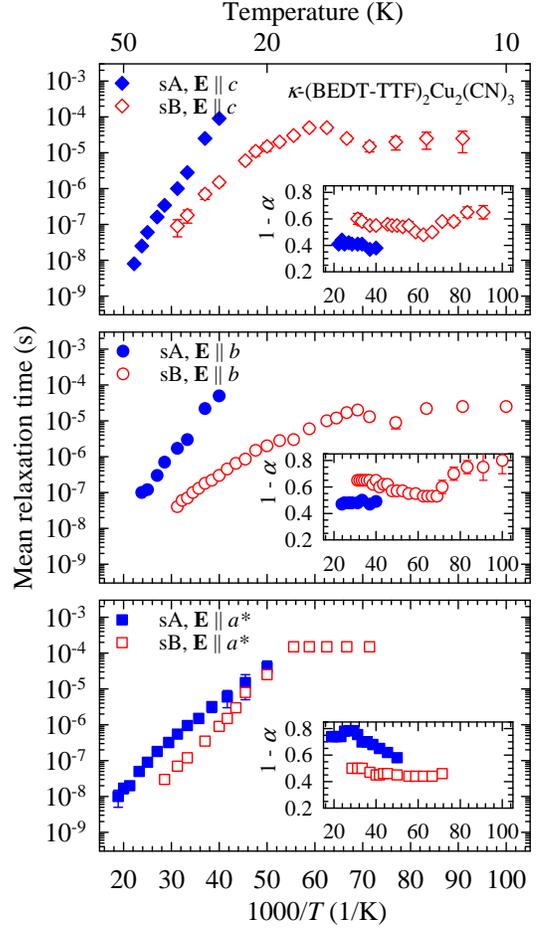} 
\caption{(Color online) Mean relaxation time and distribution of relaxation times (insets) of representative single crystals from sA (full blue symbols) and sB (open red symbols) synthesis of $\kappa$-(BE\-DT\--TTF)$_2$\-Cu$_2$(CN)$_{3}$ as a function of inverse temperature. Diamonds, circles and squares are for $\mathbf{E}\parallel c$ (upper panel), $\mathbf{E}\parallel b$ (middle panel) and $\mathbf{E}\parallel a^\ast$ (lower panel), respectively.}
\label{fig:tau3axesAB}
\end{figure}
\begin{figure}
\includegraphics[clip,width=0.8\columnwidth]{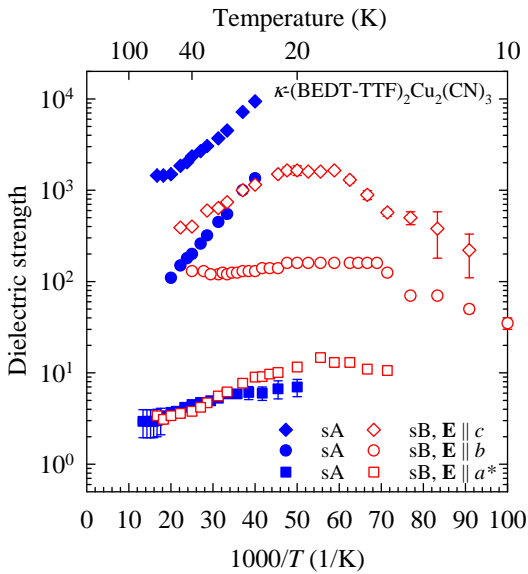} 
\caption{(Color online) Dielectric strength of representative single crystals from sA (full blue symbols) and sB (open red symbols) synthesis of $\kappa$-(BE\-DT\--TTF)$_2$\-Cu$_2$(CN)$_{3}$ as a function of inverse temperature. Diamonds, circles and squares are for $\mathbf{E}\parallel c$, $\mathbf{E}\parallel b$ and $\mathbf{E}\parallel a^\ast$, respectively.}
\label{fig:deltaepstwoaxesAB}
\end{figure}

In order to learn  how the parameters describing the relaxational process depend on disorder,
we compare the dielectric relaxation observed in single crystals from the synthesis sA and sB.
In Figs.\ \ref{fig:tau3axesAB} and \ref{fig:deltaepstwoaxesAB} the logarithms of the mean relaxation time and the dielectric strength are plotted as a function of the inverse temperature
for all three crystallographic directions.
For crystals of sA synthesis the relaxation mode shifts rapidly with decreasing temperature
and moves out of  our experimental frequency window at $T\approx 20$\,K.

\begin{figure}
\includegraphics[clip,width=\columnwidth]{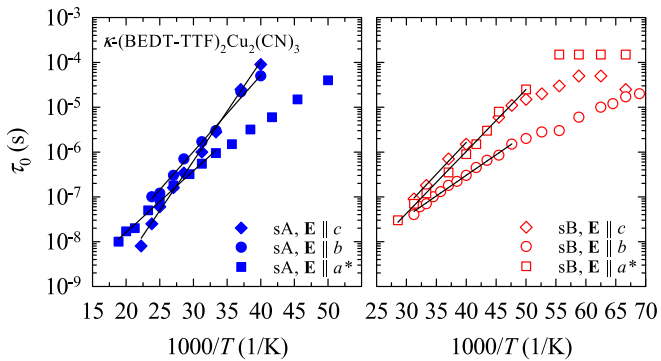} 
\caption{(Color online) Mean relaxation time in representative single crystals from sA (full blue symbols, left panel) and sB (open red symbols, right panel) synthesis of $\kappa$-(BE\-DT\--TTF)$_2$\-Cu$_2$(CN)$_{3}$ as a function of inverse temperature. Diamonds, circles and squares are for $\mathbf{E}\parallel c$, $\mathbf{E}\parallel b$ and $\mathbf{E}\parallel a^\ast$, respectively. Full lines show fits to an Arrhenius expression Eq.\ (\ref{eq:Arrhenius}) (see text).}
\label{fig:tauArrhenius}
\end{figure}

The sB samples exhibit a slower behavior of $\tau_0(T)$ upon cooling, as seen in Fig.\ \ref{fig:tau3axesAB}, and the relaxation mode can be followed down to temperatures as low as 10\,K.
Thus we can identify  a hump of the mean relaxation time and
an anomaly of dielectric constant centered around 17\,K. At lower temperatures the dielectric constant weakens within and perpendicular to planes and the relaxation time levels off.
There is an additional feature closely related to the anomalies of $\Delta\varepsilon$ and $\tau_0$
that is plotted in the insets of Fig.\ \ref{fig:tau3axesAB}:
a dip in the anomalously broad distribution of relaxation times $1-\alpha$ centered exactly in the same temperature range.
While at high temperatures the relaxation reveals an Arrhenius-type of slowing-down behavior,
the latter findings suggest that at temperature below  approximately 17\,K
the relaxation saturates and exhibits a temperature-independent relaxation time.
Hence we identify $T=17$\,K as a bifurcation temperature below which the high-temperature process freezes out, while concomitantly a low-temperature tunneling-like process sets in.\cite{Rault2000} Finally, fits to the Vogel-Fulcher formula for the sB samples basically gave similar qualitative results as for the sA crystals, as listed in Tab.\ \ref{tab:VF}. Note that the observed anomalies reduce the range of a reasonable fit by a factor of two and motivate us even more to focus on the Arrhenius behavior as the most reliable law to describe the freezing of the dielectric relaxation of \kcn{} (Fig.\ \ref{fig:tauArrhenius}, Table \ref{tab:Arhenius}).

Before proceeding further, let us consider a possible Curie-Weiss law of the
dielectric strength:
\begin{equation}
{\Delta\varepsilon} = \frac{C}{T-T_C}
\label{eq:curie}
\end{equation}
that is expected for a ferroelectric transition and discussed previously;\cite{Abdel-Jawad2010} here
$T_C$ is the Curie-Weiss temperature and $C$ is the Curie constant. If we choose $T_C$ as a free fit parameter, a satisfactory description could be obtained, as summarized in Tab.\ \ref{tab:Curie}. The negative $T_C$
observed for sB along the $b$-axis may be related to a rather weak temperature-dependent
behavior of $\Delta\varepsilon$ down to 17\,K. From the extracted values of the Curie constant
\begin{equation}
C = \frac{N p^{2}}{V \varepsilon_{0} k_B}
\end{equation}
we can estimate the charge disproportionation within the molecular dimer.
The dipole moment
\begin{equation}
p = \left({q}/{q_e}\right) d
\label{eq:dipolemoment}
\end{equation}
is given by the charge $q$ per dimer and the intradimer distance $d = 0.35$\,nm in \kcn.
Here $N$ is the number of dipole moments per unit cell volume $V$; as usual $\varepsilon_{0}$ is the permittivity of vacuum, $k_B$ is the Boltzmann constant and $q_e$ is the electron charge.
The parameters obtained from our fit to the $a^\ast$-axis data and the estimated charge disproportionation $q \approx 0.1 q_e $ agree with previous
findings of Abdel-Jawad \etal{}\cite{Abdel-Jawad2010}
Nevertheless, for all three directions parallel ($\mathbf{E}\parallel c$ and $\mathbf{E}\parallel b$) and perpendicular to the planes ($\mathbf{E}\parallel a^\ast$) the values are at least 10 to 100 times larger than the upper bound posed by optical measurements: $q/q_e \approx \pm 0.005$.\cite{Sedlmeier2012} Since electric dipoles of appreciable strength can be ruled out on the BEDT-TTF dimers, we have to look for an alternative explanation of the audio-frequency dielectric response, taking larger length scales into consideration.

\begin{table} 
\centering
\caption{Curie-Weiss parameters $C$ and $T_C$ of the dielectric strength $\Delta\varepsilon$ for representative single crystals from the sA  and sB syntheses of $\kappa$-(BE\-DT\--TTF)$_2$\-Cu$_2$(CN)$_{3}$ for the directions $\mathbf{E}\parallel c$, $\mathbf{E}\parallel b$ and $\mathbf{E}\parallel a^\ast$.}
\begin{tabular*}{\linewidth}{@{\extracolsep{\fill}}ccccc}
\hline\hline
Axis & Curie & sA & sB \\
\hline
$c$-axis & $C$ (K)		& $(4.49 \pm 0.07) \times 10^4$ K & $(1.05 \pm 0.08) \times 10^4$ K \\
$c$-axis & $T_C$ (K) 	& $20.3 \pm 0.03$ K & $16 \pm 1$ K \\
$b$-axis & $C$ (K)		& $4500 \pm 400$ K & $6500 \pm 900$ K\\
$b$-axis & $T_C$ (K) 	& $22.0 \pm 0.6$ K & $-23 \pm 7$ K \\
$a^\ast$-axis & $C$ (K)	& $213 \pm 8$ K & $99 \pm 2$ K \\
$a^\ast$-axis & $T_C$ (K) 	& $ 6 \pm 0.2$ K & $14.1 \pm 0.4$ K & \\
\hline\hline
\end{tabular*}
\label{tab:Curie}
\end{table}

The clue might be found in the quantitative differences of the dielectric response parameters
observed in crystals of different syntheses.
Most important, the in-plane mean relaxation time and dielectric strength are substantially smaller for sB samples than for the sA samples, implying that the dielectric effects become diluted as disorder increases. The difference becomes even more pronounced as the temperature decreases.
At 25\,K and $\mathbf{E}\parallel c$, for example, $\tau_0 \approx 10^{-4}$\,s and $\tau_0 \approx 10^{-6}$\,s for sA and sB crystals. The corresponding  activation energies of $\tau_0(T)$
are approximately 500 and 300\,K for sA and sB sample, respectively.
Substantial differences are also found in the magnitudes of the dielectric strength.
Again selecting $\mathbf{E}\parallel c$ at 25\,K, we find $\Delta\varepsilon \approx 10^4$ and $10^3$ for the sA and sB samples.
The observed effects suggest that disorder is more pronounced in sB crystals as compared to the sA samples, in accord with the results found in dc transport measurements.
It is interesting to note that similar effects on impurities were observed
in the dielectric response of the charge-density wave state and ferroelectric charge-ordering in quasi-one-dimensional inorganic and organic systems.\cite{Cava1985,Nad2006}
The distribution of mean relaxation times within molecular planes, however, behaves differently:
for crystals sB with higher disorder,  $1-\alpha$ is larger compared to sA crystals,
implying a lower degree of cooperativity in less homogeneous samples.
This very much resembles observations done in glassy systems.\cite{Ngai2011} An extremely broad distribution of relaxation times for dielectric relaxation within molecular planes indicates a large number of low-lying metastable energy configurations; the fact that it does not change significantly with decreasing temperature points to an already developed cooperativity in our experimental frequency and temperature window. Conversely, for sA samples and only perpendicular to molecular planes, cooperativity develops gradually as temperature decreases. This result and the result that dc transport cannot be meaningfully described by the VRH law for $\mathbf{E}\parallel a^\ast$ may be connected. 

From Fig.\ \ref{fig:VRHA} we have seen that in \kcn{} the dc conductivity $\sigma(T)$ follows the variable-range hopping law, Eq.\ (\ref{eq:VRH}), as  far as the temperature dependence is concerned.
Since in disordered systems the hopping of charges also leads to a characteristic ac response in the
radio-frequency range, we looked for a similar behavior in dielectric response observed in \kcn.
The conductivity $\sigma(\omega)$ sets in at the frequency $1/\tau_0$ roughly
proportional to the dc conductivity and is characterized by power law in the real part
with the exponent between 0.6 and 1.0.\cite{Dyre2000}
Plotting the frequency-dependent conductivity in a double logarithmic fashion ($\log \sigma$ \vs{}\ $\log \omega$), the curves taken at various temperatures should collapse
to a temperature-independent master curve.
However, in the case of \kcn{} this hopping scenario seems not to be a satisfactory description
for the data below 1\,MHz for the following reasons:
(i) In general, a power law in the ac conductivity is only observed below 1\,MHz
for disordered systems with a very high dc resistivity of $10^{10}\,\Omega$cm and higher.
In \kcn{} the in-plane and out-of-plane resistivity, however, never exceeds $10^{7}\,\Omega$cm and
$10^{9}\,\Omega$cm, respectively, even at the lowest temperature.
(ii) The onset frequency ${1/\tau_0}$ of the ac conductivity is expected to follow
the Arrhenius temperature dependence with the same activation energy as dc conductivity;
again contrary to what we observe.
(iii) The asymmetric broadening parameter $1-\alpha$ exhibits a non-monotonous temperature variation that is characterized by a pronounced dip [see Fig.\ \ref{fig:3panel3axesA}(b)],
which does not allow us to construct a master curve.

\section{Discussion}
Let us first discuss the dc conductivity and its temperature behavior.
Within the molecular planes the charge transport takes place via hopping in two dimensions.
This suggests, however, that the cation layers are a disordered system with localized states.
In fact, Kawamoto \etal{}\ first observed a remarkable broadening of the NMR line
and a variable-range hopping behavior indicating the development of inhomogeneities in the charge density upon cooling.\cite{Kawamoto2004}
Subsequent NMR results showed that the line broadening becomes strongly enhanced and anomalous below about 6\,K indicating presence of a spatially nonuniform staggered magnetization induced by an applied magnetic field. \cite{Shimizu2006} However, it is noteworthy that at very low magnetic fields,  inhomogeneity of paramagnetic origin may still persist down to very low temperatures. \cite{PrattNature2011}

Based on our comprehensive transport measurements presented above in Sec.\ \ref{sec:frequencytemperaturedependence}, we can now address this intriguing issue of Anderson localization in more detail. The localization length $\xi$ is estimated by the relation
\begin{equation} \ {k_B}{T_0} \approx \frac{\mathrm{const.}}{n_F\xi^2 d_l} \quad,
\label{eq:localization length}
\end{equation}
where $n_F$ is the density of states at the Fermi level
and $d_l$ is the interlayer spacing corresponding to the unit cell parameter $a^\ast$ perpendicular to the molecular layers; the numerical constant is in the range of $3-8$.\cite{Mott1971,NakatsujiPRL2004} Taking the VRH activation energies $T_0$ summarized in Tab.\ \ref{tab:parameters} and $n_F$ derived from specific heat data\cite{YamashitaKanoda2008,MannaPRL2010} or estimated from extended Hueckel tight-binding band calculations,\cite{Komatsu1996} we estimate $\xi$
to be smaller or at best close to the unit cell parameter.
The extremely short localization length implies
that the system is far from the metal-to-insulator transition. This seems very much in contrast
to the common phase diagram of \kcn.\cite{KurosakiPRL2005}
The linear $T$ coefficient $\gamma$ of the specific heat, however, may be attributed to the spinon density of states as first suggested by Anderson.\cite{YamashitaKanoda2008,Anderson1973}
In fact, spinons are expected to contribute to the optical conductivity, with no
significant contribution to the dc conductivity at finite temperatures.\cite{NgPRL2007,Elsaesser2012}

Thus, we conclude that the spinon density of states, associated with strong electron-electron interaction, cannot be related to Anderson-type of localization, and therefore should not be used to estimate the localization length.  We suggest that the localization in $\kappa$-CN has two distinct origins: in addition to the Mott localization due to strong electron-electron interaction, the inherent disorder provides another mechanism through the Anderson process. Recently developed theoretical approaches \cite{Byczuk2005,AguiarPRL2009} seem to be perfectly suitable to describe physics in \kcn. The strongly increasing resistivity at low temperatures may be an indication of the rather narrow bandwidth in \kcn,
because along all three crystallographic directions the intermolecular S-S contacts
exceed the van der Waals distance $d_{\text{vdW}}=0.36\,$nm.\cite{Geiser1991,Komatsu1996}
Localized states associated with Anderson localization yielding the VRH behavior might be assumed near the edges of conduction band, but we still have to clarify the origin of disorder that is crucial for variable-range hopping.

\begin{figure*}
\includegraphics*[width=14cm]{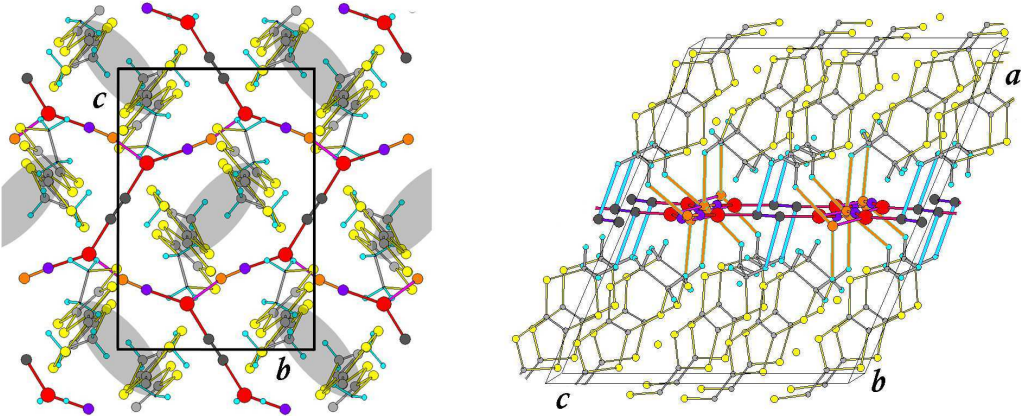}
\caption{(Color online) Left panel: View of the BEDT-TTF dimers (shaded as gray ovals) and the anion network in the $bc$ plane projected along the a-axis. Carbon, sulfur and hydrogen atoms of the BEDT-TTF molecule are colored in gray, yellow and cyan, respectively. In the anion network cooper is colored in red; carbon and nitrogen of ordered CN$^-$ groups are colored in violet and orange, while they are denoted by black in the case of CN$^-$ groups located at inversion centers. The unit cell is marked as a rectangle. Right panel: Possible hydrogen bonds between CH$_2$ groups of the BEDT-TTF molecules and CN$^-$ groups of the anion network are indicated by orange and cyan full lines; the former designate bonds to CN$^-$ groups denoted by orange and violet circles, while the latter label bonds to CN$^-$ groups located at the inversion centers.}
\label{fig:cationanionconnections}
\end{figure*}

\subsection{CN disorder}
Here we suggest that  disorder in \etcn\ originates in the Cu$_2$(CN)$_3$ sheets and indirectly affects the transport properties of the BEDT-TTF layers. As sketched in Fig.\ \ref{fig:cationanionconnections} (left panel), the anion network consists of copper ions triangularly coordinated to cyanide CN$^-$ groups.
One of them, marked by black symbols, resides on an inversion center and thus must be crystallographically disordered. Accordingly, the structural analysis based on the
high $P2_1$/c symmetry assumes a 50\% carbon and 50\% nitrogen distribution on these two atomic positions.\cite{Geiser1991,JeschkePRL2012} While the average structure with $P2_1$/c symmetry can be retained, locally the symmetry is broken. 

Structural refinements performed at 300\,K and 100\,K show that the two symmetries $P2_1$/c (with the inversion center of CN groups) and $P{2_1}$ (without the inversion center) characterize the crystal structure with comparably high agreement factors (see Table\ \ref{tab:Structure}). \cite{note3} At $T=100$\,K, where all BEDT-TTF molecules are ordered in the staggered conformation, the final $R_{\text{all}}$ (least squares residual factor for the all reflections), $wR$ (least squares weighted residual factor for reflections whose intensities are above threshold level) and GoF (least-squares goodness-of-fit) values were about 0.04, 0.05 and 1, respectively.\cite{note4} Moreover, values of the Flack index \cite{Flack1983} indicate the presence of racemic twin domains within $P{2_1}$ structure. Within the cation subsystem, formation of two dimers per unit cell and all intra and inter molecular parameters within the monoclinic unit cell, reveal no significant difference between high and low symmetry solutions. It is noteworthy that this difference is smaller than the differences between 300\,K and 100\,K found within each symmetry solution. In addition, averaged fractions of the eclipsed and staggered conformations were found equal in both symmetry solutions. 
On the other hand, there is a difference in the Cu triangular coordination within the anion network solved in the low symmetry space group: there are two different Cu triangular configurations: in the first the Cu atom is coordinated with two C atoms and one N atom, while in the second the Cu atom is coordinated with two N atoms and one C atom (see Fig.\ \ref{fig:phaseboundary}).

\begin{table} 
\centering
\caption{Structural data and refinement within space groups $P2_1$/c and $P{2_1}$ at 300\,K and 100\,K of representative single crystal from the sA synthesis of $\kappa$-(BE\-DT\--TTF)$_2$\-Cu$_2$(CN)$_{3}$. $R_{\text{all}}$, $wR$, GoF and Flack are least squares residual factor for the all reflections, least squares weighted residual factor for reflections whose intensities are above threshold level, least-squares goodness-of-fit and Flack index, respectively.}
\begin{tabular*}{\linewidth}{@{\extracolsep{\fill}}ccccc}
\hline\hline
& $P2_1$/c(300\,K) & $P{2_1}$(300\,K) & $P2_1$/c(100\,K) & $P{2_1}$(100\,K) \\
\hline
$a$ (nm) & 1.60920(4) & 1.60920(4) & 1.59644(4) & 1.59644(4) \\
$b$ (nm) & 0.85813(2) & 0.85813(2) & 0.85618(1) & 0.85618(1) \\
$c$ (nm) & 1.33904(4) & 1.33904(4) & 1.32662(3) & 1.32662(3) \\
$V$ (nm$^3$) & 1.69725(8) & 1.69725(8) & 1.65565(7) & 1.65565(7) \\
$\beta (^\circ)$ & 113.381(3) & 113.381(3) & 114.067(3) & 114.067(3) \\
$R_{\text{all}}$ & 0.0738 & 0.0775 & 0.0290 & 0.0416 \\
$wR$ & 0.0755 & 0.0787 & 0.0523 & 0.0579 \\
GoF & 1.027 & 1.020 & 1.166 & 1.027 \\
Flack & & 0.45(6) & & 0.40(5) \\
\hline\hline
\end{tabular*}
\label{tab:Structure}
\end{table}

At high temperatures fluctuations of  CN$^{-}$ entities at the inversion centers are common,
leading to variations of the angles in the Cu-C-N-Cu bonds. This is supported by density functional theory (DFT) calculations on the 4[Cu$_2$(CN)$_3$)]$^{4-}$ fragment with the fixed geometry based on the structural data. \cite{Doslic2014} Namely, the four possible orientations of these bridging CN groups, of which two are equivalent,  yielded virtually no difference in the energy, whereas modification of the CN-Cu-CN alternation along the $b$-axis considerably affected the energetics. As the temperature decreases these fluctuations become hindered and mutually contingent. Eventually the inversion symmetry is broken locally and the carbon and nitrogen atoms are considered static at one of two preferred orientation. 
Since these arrangements are not independent, local order may develop over a length scale of a few unit cells. Interestingly, the unusually broad and strong feature observed in the optical spectra at around 1\,THz [\onlinecite{ItohPRL2013}, \onlinecite{Boris}] only shows up for the electric field polarized along the direction of the disordered CN groups ($\mathbf{E}\parallel c$). This infers a coupling to the fluctuating charge which increases as the temperature is lowered down to low temperatures where it actually starts to decrease. \cite{ItohPRL2013}

The BEDT-TTF molecules are linked to the anions via hydrogen bonds: there are at least three contacts between the terminal ethylene groups and the CN$^-$ groups which are shorter than the sum of the van der Waals radii, drawn in Fig.\ \ref{fig:cationanionconnections} (right panel). Thus we can expect that the 
arrangements and fluctuations in the CN$^-$ groups will have some effect on the BEDT-TTF 
layer, for instance on the conformational degrees of freedom, i.e.\ whether the end groups 
are packed in a staggered or eclipsed fashion.\cite{PougetMCLC2006} The anions are expected to influence diverse physical properties, such as charge-density wave and charge-ordered states.\cite{Ravy1988,Foury-Leylekian2010,Alemany2012} Importantly, our DFT calculations show that the structure based on relaxed fragments is unrealistically deformed. Thus the anion-cation coupling is critical for the determination of the structural configuration of the anions in the real crystal.

\begin{figure}
\includegraphics[clip,width=\columnwidth]{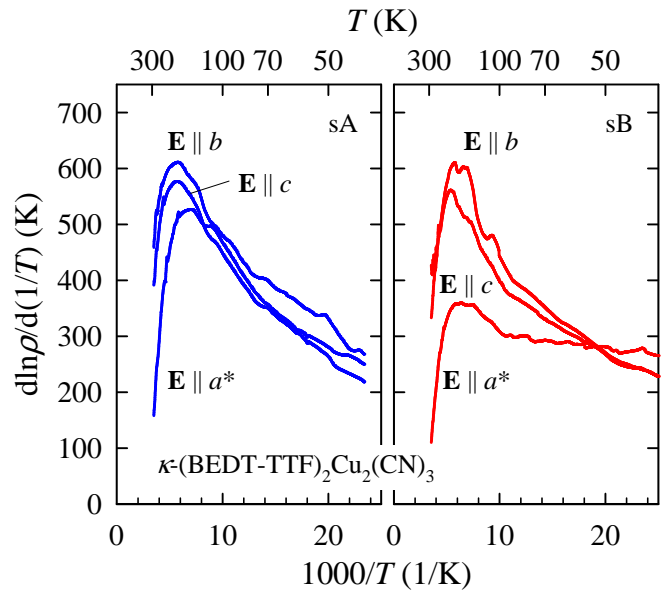}
\caption{(Color online) Logarithmic resistivity derivative for $\mathbf{E}\parallel c$, $\mathbf{E}\parallel b$ and $\mathbf{E}\parallel a^\ast$ of representative single crystals from sA (blue lines, left panel) and sB (red lines, right panel) synthesis of $\kappa$-(BE\-DT\--TTF)$_2$\-Cu$_2$(CN)$_{3}$ as a function of inverse temperature.}
\label{fig:derivatives-rho-simul}
\end{figure}

For \kcn{}  two ethylene groups out of four are disordered at room temperature and
become fully ordered in a staggered conformation around 150\,K.\cite{JeschkePRL2012} In particular, we find that for both symmetry solutions the same eclipsed and staggered fractions are present at 300 K, while only the staggered one is present at 100\,K. 
We also see a rapid change in the temperature-dependent resistivity in this temperature range,
as demonstrated in Fig.\ \ref{fig:derivatives-rho-simul} where the derivative of $\rho(T)$
is plotted for the in-plane and out-of-plane direction.
The ordering develops gradually between $T=200$ and 120\,K, right in the temperature range where the nearest-neighbor dc hopping crosses over into the variable-range hopping.
Most remarkable, temperature and width seem to depend on sample synthesis and
thus evidence the effect of disorder.

We conclude, that the dc transport that takes place in the BEDT-TTF layers
is governed by the random potential created by the CN$^-$ groups on the inversion centers.
This charged disorder is rather well screened at high temperatures and
becomes gradually more pertinent at low temperatures.
As the temperature decreases below $T_c$, hops over longer distance but closer in energy become more favorable promoting the VRH law. Finally, our data also show that the samples grown under different conditions yield different values of the $T_0$ VRH parameter. This result indicates the presence of a different degree of the structural disorder in the anion subsystem.

It is noteworthy that the x-ray irradiation effects on dc transport were recently observed and attributed to molecular defects resulting in the effective carrier doping into the half-filled band.\cite{Sasaki2007} High enough irradiation level was found to induce a metallic-like dc transport above 200\,K, while insulating behavior but with smaller resistivity and activation energy was found at lower temperatures. This result is not surprising since we expect that irradiation-induced defects, mainly created in the anion layers \cite{Sasaki2012}, should enhance the inherent random potential present in nominally pure single crystals as discussed above.

\subsection{Dielectric response}
\begin{figure}
\includegraphics[clip,width=0.8\columnwidth]{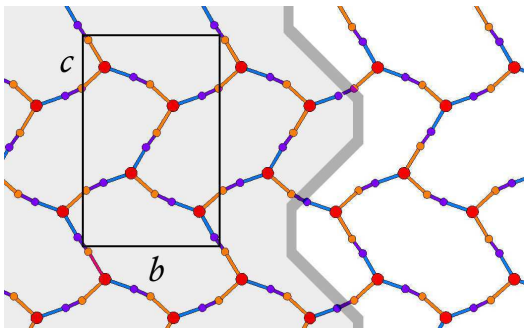}
\caption{(Color online) View of the anion network in the $bc$ plane projected along the $a$-axis tilted by 24 deg. The unit cell is marked as a rectangle. Assumed local inversion symmetry breaking promotes formation of an anti-phase boundary between two neighboring domains. Gray thick line denote the phase boundary. Note that the 50\% carbon - 50\% nitrogen occupancy of the original average structure is preserved.}
\label{fig:phaseboundary}
\end{figure}

With these observations in mind we now turn to the second intriguing issue of \kcn{}: 
the presence of a dielectric response despite no electric dipoles are associated with the BEDT-TTF dimers as established by infrared vibrational spectroscopy.\cite{Sedlmeier2012} The latter data discard the original interpretation by Abdel-Jawad and Hotta \cite{Abdel-Jawad2010,Hotta} based on the collective excitation of the intradimer electric dipole; this idea was previously also used to explain the microwave and terahertz anomalous charge behavior.\cite{Poirier2012,ItohPRL2013}  Nevertheless, the vibrational data allow for the presence of fast temporal fluctuations of charge distribution with the exchange frequency of $10^{11}$ Hz, whose softening is predicted theoretically. \cite{Naka} While fluctuating intradimer electric dipoles have been still discussed in literature as the possible source of terahertz response\cite{ItohPRB2013}, it is definitely clear that they cannot be invoked to explain dielectric response at low frequencies. Thus, we propose that the disorder in the CN$^-$ groups acts on the BEDT-TTF layers via the hydrogen bonds,
where it alters the charge distribution and causes domain boundaries, 
as depicted in Fig.\ \ref{fig:phaseboundary}.

We suggest the presence of line and/or point defects within phase boundaries separating domains with structurally ordered CN$^-$ groups. This potential is mapped onto the BEDT-TTF layer via the hydrogen bonds. These boundaries and defects carry charge which respond to an applied ac field. Thus over a wide temperature range there is a mixture of domains of various sizes and configurations of ordered CN$^-$ groups with balance becoming gradually such that the number of interfaces and phase boundaries increases down to about 17\,K. Concomitantly, the system takes longer and longer time to respond indicating an increasingly cooperative character which requires larger and larger activation energies to achieve local rearrangements.
The latter is expected as the size of cooperatively relaxing regions grows with decreasing temperature concomitantly resulting in a decreasing weight of the interfaces. Keeping in mind that the dielectric strength is proportional to the density of collective excitations, we thus conclude that the number of charged defects increases down to about 17\,K and then decreases again. 
It is noteworthy that the change in the temperature behavior of $\Delta\varepsilon$  coincides with changes observed in the temperature dependence of the structural parameters: The steady trend persisting from high temperatures levels off at these low temperatures.\cite{JeschkePRL2012} 
Our low-frequency dielectric data reveal fingerprints typical for relaxor ferroelectricity whose dynamics is characterized by gradual slowing down and tunneling at high and low temperatures, respectively. Glassy freezing of this relaxation happens between 7\,K and 15\,K, slightly above 6\,K were strong lattice effects were observed. \cite{MannaPRL2010} We suspect that the glassiness results from constraints on the effective dynamics of the \kcn{} system as a complex cation-anion system intrinsically coupled by hydrogen bonds. Such a complex nature of \kcn{}  gives rise to an exotic charge-spin-lattice coupling yielding a mutually interconnected anomalous behavior of physical properties in charge, spin and lattice sectors.\cite{Kawamoto2004,MannaPRL2010,Poirier2012,Poirier2013,ItohPRL2013, ItohPRB2013} We want to recall the suggestion \cite{Shimizu2006} that the spatially inhomogeneous magnetization with anomalous features originates from symmetry-broken sites. This might also reflect the structural properties of the anion network. \cite{Gregor2009}

A couple of additional points are worth to stress. In the suggested picture, the heterogeneous character of the dielectric relaxation is a consequence of a structural property of the anion network, which induces the break-up of the system in a mosaic of presumably frustration-limited domains. In addition, the formation of the magnetic ordering may not be solely prohibited
due to triangular arrangement of BEDT-TTF dimers; rather it may be a combined effect of frustration and proposed symmetry breaking of the anion network.\cite{QiPRB2008} As a matter of fact, doubts have been raised whether the quantum spin liquid phase in \kcn{} is solely caused by geometrical frustration; statistical disorder might also be decisive.

\section{Summary}
Our investigations of the frequency and temperature dependent complex conductivity yield clear evidence for variable-range hopping transport within the molecular planes in dc limit and an anomalously broad anisotropic dielectric relaxation in the audio-frequency range below 60\,K. The relaxation bears typical fingerprints of relaxor ferroelectricity: a gradual slowing down becomes dominated by tunneling at low temperatures and at the bifurcation temperature we detect anomalous features in dielectric constant, mean relaxation time and its distribution. These results demonstrate the inherent heterogeneity present in nominally pure single crystals of \etcn{}. We argue that the observed effects can be accounted for by charge defects generated in interfaces between frustration-limited domains whose creation is triggered by an inversion-symmetry breaking at a local scale in the anion network. Experiments such as x-ray diffuse scattering and x-ray diffraction at low temperatures combined with first-principles density functional calculations are envisaged as desirable tools to elucidate our proposal. In addition, novel theoretical efforts within Mott-Anderson approach are needed in order to refine microscopic understanding of charge-spin-lattice coupling in the presence of disorder and the formation of spin-liquid state in \etcn{}.

\begin{acknowledgments}
Technical assistance in x-ray diffraction measurements of D.\ Matkovi\'{c}-\v{C}alogovi\'{c} and ab-initio calculations done by N.\ Do\v{s}li\'{c} are gratefully acknowledged. We thank K.\ Biljakovi\'{c}, S.\ Ishihara, J.\ P.\ Pouget, G.\ Saito and D.\ Stare\v{s}ini\'{c} for very helpful discussions. The work has been supported by the Croatian Science Foundation project IP-2013-11-1011 and the DAAD German-Croatian project. We appreciate financial support by the Deutsche Forschungsgemeinschaft (DFG).
\end{acknowledgments}


\begin{thebibliography}{99}
\bibitem{note1}
The samples sA and sB were synthesized using the following electrocrystallization route: a solution of ET and CuCN, KCN, 18-Crown-6 was used in a solvent containing mixture of a 1,1,2-trichloroethane:ethanol in ratio 90:10 with growth period of 3 and 10 months and constant current of 1\,$\mu$A and 0.25\,$\mu$A, respectively. For samples sC electrolyte mixture was the same, but for solvent a mixture of  chloroform and ethanol in ratio 90:10, and a drop of water was used. These crystals were grown for 42 days using constant current of 0.25\,$\mu$A.
\bibitem{note2}
The dielectric relaxation can be detected only at temperatures below 60\,K.
Above $T\approx 50$\,K we can determine only the dielectric relaxation strength by measuring the capacitance at 1\,MHz.
\bibitem{note4}
Residual factor for all reflections is 
$R_{\text{all}} = \sum_{\text{all}} ||{F_o}| - |{F_c}|| / \sum_{\text{all}}|{F_o}|$; 
weighted residual factor for reflections whose intensities are above threshold level 
$wR = [\sum_{F>4\sigma(F)} w(F_o^2 - F_c^2)^2 / \sum w(F_o^2)^2]^{1/2}$; goodness-of-fit 
$\text{GoF} = [\sum w(F_o^2 - F_c^2)^2 / (n - p) ]^{1/2}$ taking into account number $n$ of reflections and total number $p$ of refined parameters. Flack index is calculated in terms of absolute structure forming one out of two domains of racemic twins.
\end{thebibliography}
\end{document}